\documentclass[journal,twocolumn]{IEEEtran}
\IEEEoverridecommandlockouts

\usepackage{cite}
\usepackage{amsmath,amssymb,amsfonts}
\usepackage{algorithmic}
\usepackage{graphicx}
\usepackage{textcomp}
\usepackage{xcolor}
\usepackage{flushend}
\usepackage{tikz}
\usepackage{subfig}
\usetikzlibrary{shapes, arrows}
\def\BibTeX{{\rm B\kern-.05em{\sc i\kern-.025em b}\kern-.08em
    T\kern-.1667em\lower.7ex\hbox{E}\kern-.125emX}}
    
\IEEEaftertitletext{\vspace{-1.0\baselineskip}}
\begin{document}

\title{Coded Faster-than-Nyquist Signaling for Short Packet Communications}

\author{
{Emre Cerci$^1$, Adem Cicek$^1$, Enver Cavus$^1$, Ebrahim Bedeer$^2$, Halim Yanikomeroglu$^3$}\\
\IEEEauthorblockA{$^1$Electrical and Electronic Eng. Department, Ankara Yildirim Beyazit University, Ankara, Turkey.\\
$^2$Department of Electrical and Computer Engineering, University of Saskatchewan, Saskatoon, SK, Canada.\\
$^3$Department of Systems and Computer Engineering, Carleton University, Ottawa, ON, Canada. \\
E-mails: emre-cerci@hotmail.com, \{acicek, ecavus\}@ybu.edu.tr, e.bedeer@usask.ca, halim@sce.carleton.ca.
}

}

\maketitle

\begin{abstract}
Ultra-reliable low-latency communication (URLLC) requires short packets of data transmission. It is known that when the packet length becomes short, the achievable rate is subject to a penalty when compared to the channel capacity. In this paper, we propose to use faster-than-Nyquist (FTN) signaling to compensate for the achievable rate loss of short packet communications. We investigate the performance of a combination of a low complexity detector of FTN signaling used with nonbinary low-density parity-check (NB-LDPC) codes that is suitable for low-latency and short block length requirements of URLLC systems. Our investigation shows that such combination of low-complexity FTN signaling detection and NB-LDPC codes outperforms the use of close-to-optimal FTN signaling detectors with LDPC codes in terms of error rate performance and also has a considerably lower computational complexity.
\end{abstract}


\section{Introduction}
\tikzstyle{block} = [draw, rectangle, fill=black!20, text width=6em, text centered, minimum height=13mm, node distance=15mm] 
\tikzstyle{inv} = [minimum height=13mm, text width=0em, node distance=15mm] 
\tikzstyle{line} = [draw, -stealth, thick] 									
\begin{figure*}[t]
\centering
\begin{tikzpicture}
\node [inv,   					 xshift=-40] (invstart) { };
\node [block, right of=invstart, xshift=40]  (start) 	{Channel Encoder};
\node [block, right of=start, 	 xshift=40]  (mod)   	{Modulator};
\node [inv,   right of=mod, 	 xshift=40]  (binv)  	{ };
\node [block, right of=binv, 	 xshift=110]  (trf)   	{Transmit Filter};
\node [block, right of=trf, 	 xshift=40, minimum height = 12mm, yshift=-36] (channel) {AWGN Channel};
\node [block, below of=trf, 	 yshift=-25] (mf) 		{Matched Filter};
\node [block, below of=binv, 	 yshift=-25] (se) 		{FTN Detector};
\node [block, below of=mod, 	 yshift=-25] (demod) 	{Demodulator};
\node [block, below of=start, 	 yshift=-25] (end) 		{Channel Decoder};
\node [inv,   below of=invstart, yshift=-25] (invend) 	{};
\path [line] (start) 	-- node [yshift=10]  				{$\pmb{c}$} (mod);
\path [line] (mod) 		-- node [yshift=10]  				{$\pmb{a}$} (trf);
\path [line] (trf) 		-| node [yshift=10,  xshift=-25]  	{$s(t)$} (channel);
\path [line] (channel) 	|- node [yshift=-10, xshift=-25] 	{$r(t)$} (mf);
\path [line] (mf) 		-- node [yshift=10,  xshift=30]  	{$y(t)$} (se);
\path [line] (mf) 		-- node [yshift=10,  xshift=-30]    {$y_k$} (se);
\path [line] (mf) 		-- node [yshift=-10]    			{Sampling at $\tau T_0$} (se);
\path [line] (se) 		-- node [yshift=10]  				{$\hat{\pmb{a}}$} (demod);
\path [line] (demod) 	-- node [yshift=10]  				{$\hat{\pmb{c}}$} (end);
\path [line] (invstart) -- node [yshift=10]  				{$\pmb{k}$} (start);
\path [line] (end) 		-- node [yshift=10]  				{$\hat{\pmb{k}}$} (invend);
\path [line] (invstart) -- node [yshift=-10]  				{Input} (start);
\path [line] (end) 		-- node [yshift=-10]  				{Output} (invend);
\fill[white] (9.7,-2.5) rectangle (10.25,-2.3);			
\draw[thick](10.25,-2.38) -- (9.8, -2);					
\draw[dashed, thick] (9.8, -2) arc (139.82:180:.59);	
\end{tikzpicture}
\caption{Block diagram of a coded FTN signaling system.}
\label{blockdiagram}
\end{figure*}
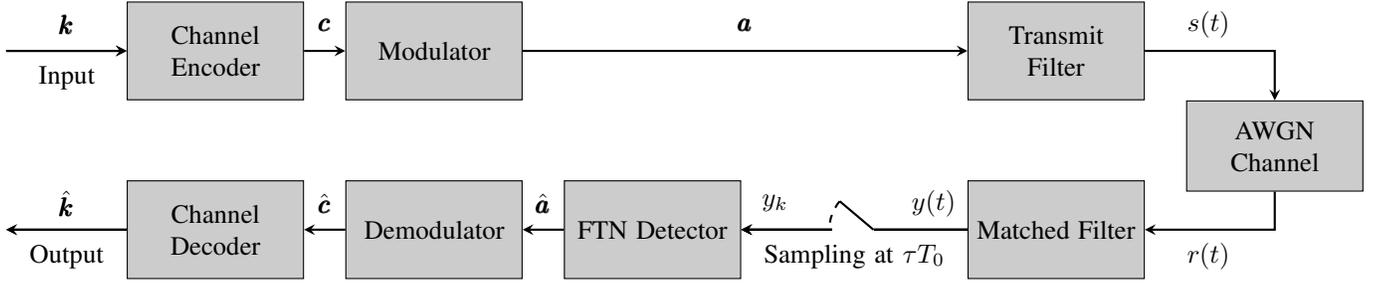

Ultra-reliable low-latency communications (URLLC) improve reliability and latency for next-generation wireless communication systems. URLLC communications require short packets of information to be transmitted and this causes loss of the achievable rate when compared to the channel capacity \cite{Polyanskiy}. Recently, faster-than-Nyquist (FTN) signaling was proposed as a candidate solution to compensate for the achievable rate loss due to the short packet length \cite{rate0}. 

In Nyquist signaling, the use of orthogonal pulses for transmission avoids inter-symbol interference (ISI), yet decreases the spectral efficiency (SE). In 1975, J. E. Mazo \cite{mazo1} showed that sinc pulses carrying binary information bits can be accelerated up to 25\% beyond the Nyquist rate, and hence, violates the orthogonality condition, without affecting the asymptotic error rate. That said, FTN signaling uses nonorthogonal pulses for transmission to make a better utilization of the bandwidth in return of controlled ISI. The existence of ISI between neighboring pulses can cause deterioration of the detection performance if proper sequence estimation techniques are not in place at the receiver. Recently, several FTN signaling detectors have been proposed to balance the detection performance and the signal processing complexity at the receiver \cite{bcjr4, gbk6, mbcjr5, SDR, GS}.

Channel codes in conventional communications systems assume infinite block length and when used for finite block length transmission, their achievable rate will have a penalty when compared to the channel capacity. For instance, for very long codeword lengths, low-density parity-check (LDPC) codes with iterative Belief Propagation (BP) decoding \cite{mackayldpc9} perform very close to the Shannon limit; however, they do not maintain the same superior performance for finite block lengths. As an extension of LDPC codes, nonbinary-LDPC (NB-LDPC) codes are efficient codes in terms of correcting channel errors for finite block lengths \cite{nbvsb7}. In particular, their performances were observed to be better on high order Galois Fields, $GF(q)$, than their LDPC counterparts \cite{mackaygfq10} and promising BER performances were observed for finite block lengths \cite{nbvsb7}. The decoding of NB-LDPC codes can be done with the extended min-sum (EMS) decoder \cite{nbldpc11} that can decrease the decoding complexity by considering a limited number of the available likelihoods for each message of the Tanner graph instead of considering all the available $q$ likelihoods for calculations.

In this work, we investigate the performance-complexity trade-offs when using FTN signaling for short packet communications.
In particular, for FTN signaling, we consider two extreme detectors, i.e., successive symbol-by-symbol with go-back-\textit{K} sequence estimation (SSSgb\textit{K}SE) \cite{gbk6} and M-algorithm Bahl-Cocke-Jelinek-Raviv (M-BCJR) \cite{bcjr4} as the lowest known complexity FTN signaling detector and the best performing FTN signaling detector, respectively. 
With its superior performance, M-BCJR is a close-to-optimal detector, but also has high complexity.
We also consider LDPC codes of block lengths of 128, 256, and 512 bits and NB-LDPC codes of block lengths of 120, 264, and 504 bits. Our investigation shows that for a time acceleration of $\tau=0.7$, NB-LDPC with SSSgb\textit{K}SE FTN signaling detection does not outperform LDPC with M-BCJR FTN signaling detection. However, for $\tau=0.8$, NB-LDPC with SSSgb\textit{K}SE FTN signaling detection has better performance than LDPC with M-BCJR FTN signaling detection for low signal-to-noise ratios (SNRs). For higher values of $\tau=0.9$, NB-LDPC with SSSgb\textit{K}SE FTN signaling detection always outperforms LDPC with M-BCJR FTN signaling detection, yielding promising SNR gains.
Such results suggest that for light ISI scenarios, it is sufficient to use a low-complexity FTN signaling detector with NB-LDPC rather than using a complex FTN signaling detector with LDPC. Our complexity analysis reveals that the combination of NB-LDPC with SSSgb\textit{K}SE FTN signaling detection is of considerably lower complexity when compared to LDPC with M-BCJR FTN signaling detection.

The remainder of this paper is organized as follows. Section II introduces the coded FTN signaling system model. In Section III, we discuss the architectures of the channel coding and FTN signaling detector used in this study. Section IV gives the simulation results, while Section V provides the complexity comparisons. The paper is finally concluded in Section VI.

\section{System Model}
The general block diagram of a coded FTN signaling communication system is shown in Fig.~\ref{blockdiagram}. At the transmitter, an input message $\pmb{k}$ is encoded using the channel encoder block and then the encoded output, $\pmb{c}$, is converted to symbols, $\pmb{a}$, in the modulator block. Finally, the modulated symbols are passed through a root-raised cosine (rRC) pulse, where FTN signaling is applied. In transmit pulse, the modulated symbols are first shaped by a unit energy pulse, $p(t)$, i.e., $\int_{-\infty}^{\infty} |p(t)|^2 dt = 1$. Then, each symbol is transmitted every $\tau T_0$, where $\tau$ is the time acceleration/packing parameter and $T_0$ is the symbol duration. The value of $\tau$ is bounded by 0 and 1, i.e., $0<\tau\leq1$, and $\tau=1$ means the conventional Nyquist signaling. The transmit signal $s(t)$ can be written as
\begin{equation}
s(t)=\sqrt{E_s} \sum_{n=1}^{N} a[n]p(t-n\tau T_0),\label{eq1}
\end{equation}
\noindent
where $E_s$ is the energy of the data symbol $a[n]$ drawn from a binary phase shift keying (BPSK) and {$N$ is the total number of transmitted symbols}. The transmit signal is sent through an additive white Gaussian noise (AWGN) channel, and the received signal $r(t)$ can be expressed as
\begin{equation}
r(t)=\sqrt{E_s} \sum_{n=1}^{N} a[n]p(t-n\tau T_0)+n(t),\label{eq2}
\end{equation}
\noindent
where $n(t)$ is a Gaussian noise with zero mean and a variance of $\sigma^{2}$. 
At the receiver, assuming a matched filter to $p(t)$, the signal $y(t)$ after the matched filter is expressed as
\begin{equation}
y(t)=\sqrt{E_s} \sum_{n=1}^{N} a[n]g(t-n\tau T_0)+w(t),\label{eq3}
\end{equation}
\noindent
where $g(t)=\int_{-\infty}^{\infty} p(z)p(z-t) dz$ and $w(t)=\int_{-\infty}^{\infty} n(z)p(z-t) dz$ represent the additive effect of noise after filtering. At the output of the matched filter, $y(t)$ is sampled at every $\tau T_0$ and the sampled signal $y_k$ can be expressed as
\begin{align}
y_k=\sqrt{E_s} a_k g(0) + \sum_{n=1,n\neq k}^{N} a[n]g((k-n)\tau T_0)+w(k\tau T_0),\label{eq4}
\end{align}
\noindent
where $\sqrt{E_s} a_k g(0)$ is the signal sample having the information of $a_k$, $w(k\tau T_0)$ is the sampled noise value, and $\sum_{n=1,n\neq k}^{N} a[n]g((k-n)\tau T_0)$ is the ISI from adjacent symbols. Then, the sampled signal is passed to the FTN signaling detector block to obtain the symbol estimates, $\hat{\pmb{a}}$. Finally, the demodulator generates soft log-likelihood ratio (LLR) values per bit, which are then used by the channel decoder to estimate $\hat{\pmb{k}}$ for the transmitted bit sequence $\pmb{k}$.

\section{NB-LDPC Coded Short Packet FTN Communication System with Symbol-by-Symbol Detection}
In this section, we study a NB-LDPC coded short packet FTN signaling communication system with SSSgb\textit{K}SE as the FTN signaling detector. It is known that the BCJR algorithm or its variants provide the optimum or near optimum FTN detection performance; however, their detection complexity and latency are not suitable for short packet communications. Therefore, in this section, we adopt SSSgb\textit{K}SE \cite{gbk6} to be used in short packet FTN signaling communications. Then, we try to compensate for the performance loss introduced by the low-complexity detector using NB-LDPC codes, which are known to achieve good performance even at finite block lengths.
\subsection{Symbol-by-Symbol with go-back-\textit{K} Sequence Estimator}\label{sub_detector}
As it is observed from \eqref{eq4}, the ISI caused by adjacent symbols can be removed by a simple subtraction if the interference from adjacent symbols can be estimated. For a noise-free transmission, the relation between the received sample vector $\pmb{y}$ and the transmit data symbols vector $\pmb{a}$ is $\pmb{y}=\pmb{Ga}$, where the ISI matrix $\pmb{G}$ has the elements $G_{i,j}=g((i-j)\tau T_0)$ that refer to the ISI between $i$th and $j$th data symbols \cite{gbk6}. Assuming $L -1$ is the length of ISI from one side of a symbol, there is $2(L-1)$ ISI values affecting a single symbol. Consequently, a signal sample $a_k$ can be recovered from $y_k$ by removing the effect of the ISI from previous and upcoming symbols with respect to the index $k$ as in (5):
\begin{align}
\label{eq5}
G_{1,1} a_k=y_k &-(G_{1,L} a_{k-L+1}+...+G_{1,2} a_{k-1}) \nonumber\\
&-(G_{1,2} a_{a+1}+...+G_{1,L} a_{k+L-1} ).
\end{align}
However, at any instance of a serial transmission, none of upcoming symbols is received even though the ISI is present. Thus, the ISI is partially removed only from the previous estimated symbols in \eqref{eq5} and $a_k$ can be estimated as
\begin{equation}
\hat{a}_k={\textup{quantize}}\{y_k-(G_{1,L} a_{k-L+1}+...+G_{1,2} a_{k-1} )\},\label{eq6}
\end{equation}
\noindent
where $\hat{a}_k$ is the $k$th estimated symbol and the function ${\textup{quantize}}\{x\}$ maps $x$ to the nearest modulation symbol. This method is called as successive symbol-by-symbol sequence estimation (SSSSE) in \cite{gbk6}. Once $\hat{a}_k$ is estimated, the value of $\hat{a}_k$ can be used to remove its ISI effect on the previous symbols. In other words, after the determination of $\hat{a}_k$, the previous symbols can be re-estimated by removing the ISI due to $\hat{a}_k$. This technique is the main idea behind SSSgb\textit{K}SE and $(k-K)$th symbol is re-estimated as follows
\begin{align}
\label{eq7}
\hat{\hat{a}}_{k-K}=&\quad{\textup{quantize}}\bigg\{y_k\nonumber\\
&- (G_{1,L} \hat{a}_{k-K-L+1}+...+G_{1,2} \hat{a}_{k-K-1})\nonumber\\
&- (G_{1,2} \hat{\hat{a}}_{k-K+1}+...+G_{1,K+1} \hat{\hat{a}}_{k})\bigg\},
\end{align}
\noindent
where $\hat{\hat{a}}_j$ stands for the re-estimated $j$th symbol while $\hat{a}_j$ is the estimated signal with SSSSE. If \eqref{eq7} is continued until $(k-1)$th symbol, accuracy of previous \textit{K} symbols is enhanced. After the symbol sequence estimation with SSSgb\textit{K}SE, $\hat{\pmb{a}}$ is demodulated and soft LLR values are generated for the channel decoder. 

\subsection{NB-LDPC Codes}\label{sub_ecc}

NB-LDPC codes are an extension of LDPC codes to a Galois Field $GF(q)$ with $q>2$ and they have been observed to provide good BER performance at short codeword lengths \cite{nbvsb7}. The parity check matrix (PCM) $\pmb{H}$ of NB-LDPC is still very sparse, yet the elements belong to $GF(q)$. An NB-LDPC code can be downgraded to an LDPC code with a PCM of size $mM_H\times mN_H$ for $q=2^m$. Any full location of an NB-LDPC PCM refers to a cyclic shift applied on an $m\times m$ identity matrix, $\pmb{I}_m$, depending on the element from $GF(q)$. Compared to LDPC error correction techniques, using finite field arithmetic to correct errors is more complex, but more reliable.

An NB-LDPC encoder takes a $K_H$-dimensional input message vector $\pmb{k}$ with elements from $GF(q)$, where $K_H=M_H-N_H$. An $M_H$-dimensional parity vector $\pmb{p}$ is found such that $\pmb{c}=\big[\begin{matrix}\pmb{p} & \pmb{k}\end{matrix}\big]$ satisfies all parity check equations expressed in \eqref{eq8}:
\begin{align}
f_i=\sum_{j=1}^{N_H} H_{i,j}c_j&,& H_{i,j} \in GF(q),\label{eq8}
\end{align}
\noindent
where $f_i$ is the $i$th parity check equation, and the multiplication is in $GF(q)$. $f_i$ is said to be satisfied as long as $f_i=0$ holds.

The encoded codeword $\pmb{c}$ is modulated with BPSK bits-to-symbols mapping. As BPSK takes binary inputs, $\pmb{H}$ has to have elements from a finite field $GF(2^m)$ such that each codeword element has $m$ bits. Consequently, the modulated symbol sequence $\pmb{a}$ contains $m N_H$ symbols.

Even though NB-LDPC codes provide good performance at short block lengths, the BP decoding of NB-LDPC codes has very high complexity. The EMS algorithm \cite{voicila12} is introduced as a low complexity decoder for NB-LDPC codes and it is more suitable for short packet communication \cite{songlin13}. The EMS decoder works in the log-domain using only $n_m$ most trusted probabilities in each iteration, where $n_m\ll q$. The EMS algorithm reduces the complexity of $\mathcal{O}(q^2)$ for the BP decoder to $\mathcal{O}(n_m \log_2 n_m)$ \cite{voicila12} and it results in a faster decoder. 

EMS decoder is a log-likelihood ratio (LLR) based decoder. A LLR vector for a codeword symbol is $\pmb{\ell}(z)=\begin{bmatrix} \ell_0 & \ell_1 & ... & \ell_{q-1} \end{bmatrix}^T$ and the likelihood of a message $\alpha_i$ can be expressed as
\begin{align}
\ell_i=\log \frac{P(z=\alpha_i)}{P(z=\alpha_0)}&,& i\in\{0, ..., q-1\}, \label{eq9}
\end{align}
\noindent
for a random variable $z$ representing a connection in the Tanner graph and $\alpha_i\in GF(q)$ denotes the message on the connection. The notion \textit{message} in a decoder stands for the likelihoods of a single element. NB-LDPC codes are based on iterative decoders and the EMS decoder halts if all syndromes are satisfied or a predetermined number of iterations are reached. A single iteration of the decoder consists of finding variable-to-check (V2C) messages, check-to-variable (C2V) messages and syndrome control. The key difference between LDPC and NB-LDPC is that NB-LDPC messages have more than two LLR values and the messages are represented in multidimensional vectors having $q$ elements.

A NB-LDPC C2V operation is based on finding the input configurations that satisfy corresponding check node. This results in a complexity of $\mathcal{O}(q^2)$ for BP algorithm. Handling convolutions in Fourier domain reduces this complexity to $\mathcal{O}(q\log_2 q)$ \cite{fastnbldpc14}. The EMS algorithm uses elementary check nodes (ECN) to simplify check node inputs and truncated messages to reduce $q$-dimensional LLR to the most useful $n_m$ values. Using $n_m\ll q$ still results in acceptable error performance and comparably lower complexity of $\mathcal{O}(n_m \log_2 n_m)$. The omitted $q-n_m$ values are filled with a scalar value \cite{voicila12}, $\gamma_A$ , having the expression
\begin{equation}
\gamma_A = \max_{i=0, A[i]\notin\pmb{B}} (A[i]) - \log{(q-n_m)} - \textup{offset}, \label{eq12}
\end{equation}
\noindent
where $\pmb{A}$ and $\pmb{B}$ are regular and truncated LLR vectors, respectively.

In a V2C operation, variable-to-check message vector, namely $\pmb{R}$, is found as follows
\begin{equation}
R_{d_v}[\alpha_k] = I_{llr}[\alpha_k]+\sum_{i=1}^{d_v -1} Q_i[\alpha_k],\quad \begin{aligned} \alpha_k\in GF(q),\end{aligned} \label{eq13}
\end{equation}
where $d_v$ is the number of connections to a variable node in Tanner graph, $\pmb{I}_{llr}$ is input LLR vector, and $\pmb{Q}$ is C2V message vector. At the beginning of decoding process, $\pmb{R}$ and $\pmb{Q}$ are initialized as zero. For the truncated vector $\pmb{Q}_i$ in \eqref{eq13}, $Q_i [\alpha_k ]=\gamma_{Q_i}$ is used if $\alpha_k\notin \pmb{\beta}_{\pmb{Q}_i}$, where $\pmb{\beta}_{\pmb{Q}_i}$ contains locations of $n_m$ values in $GF(q)$ after truncation.

\begin{figure*}[t!]
	\centering
	\subfloat[\label{ber107}]{
		\includegraphics[width=0.3\textwidth]{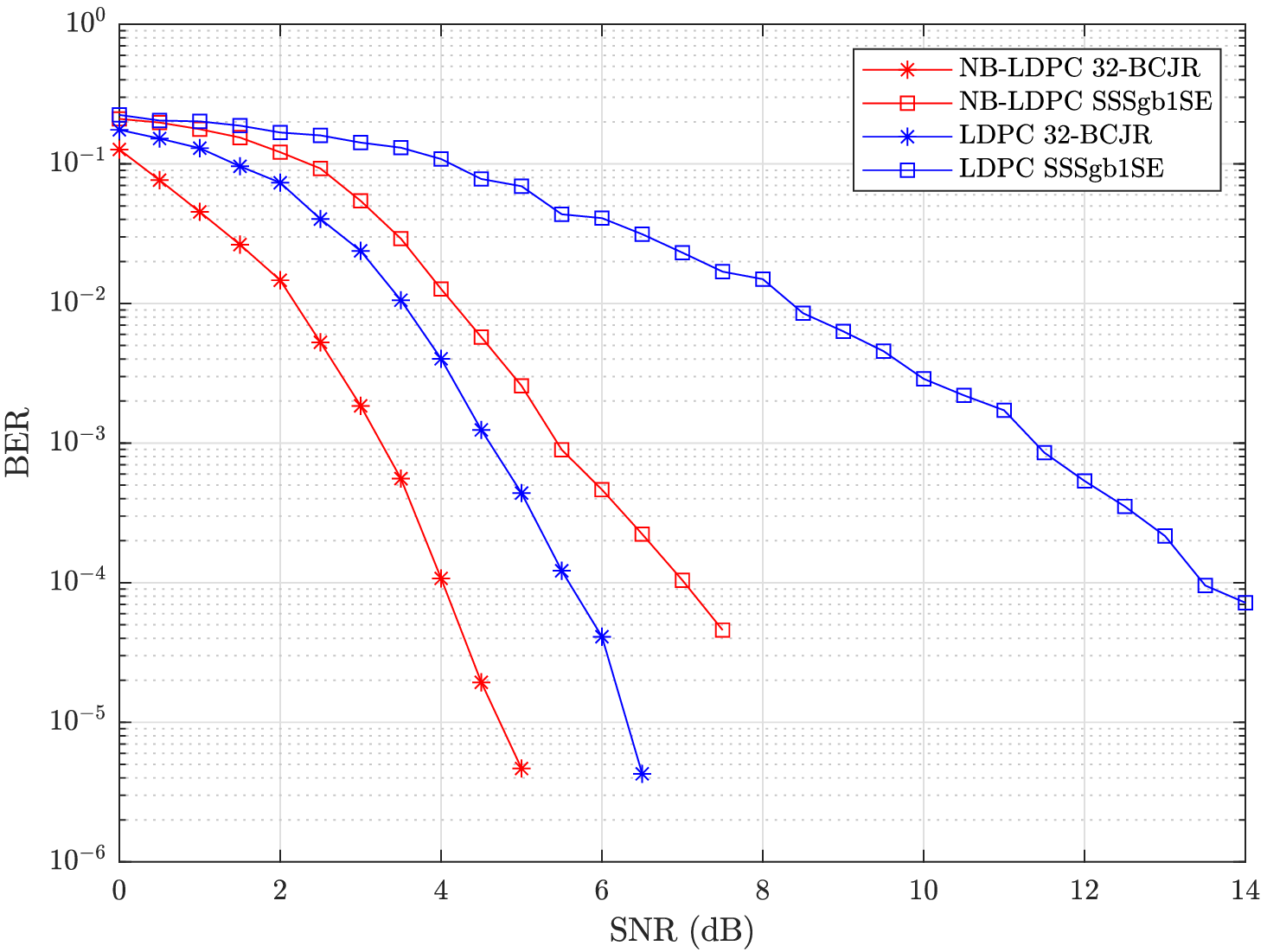}}
		\hfill
	\subfloat[\label{ber207}]{
		\includegraphics[width=0.3\textwidth]{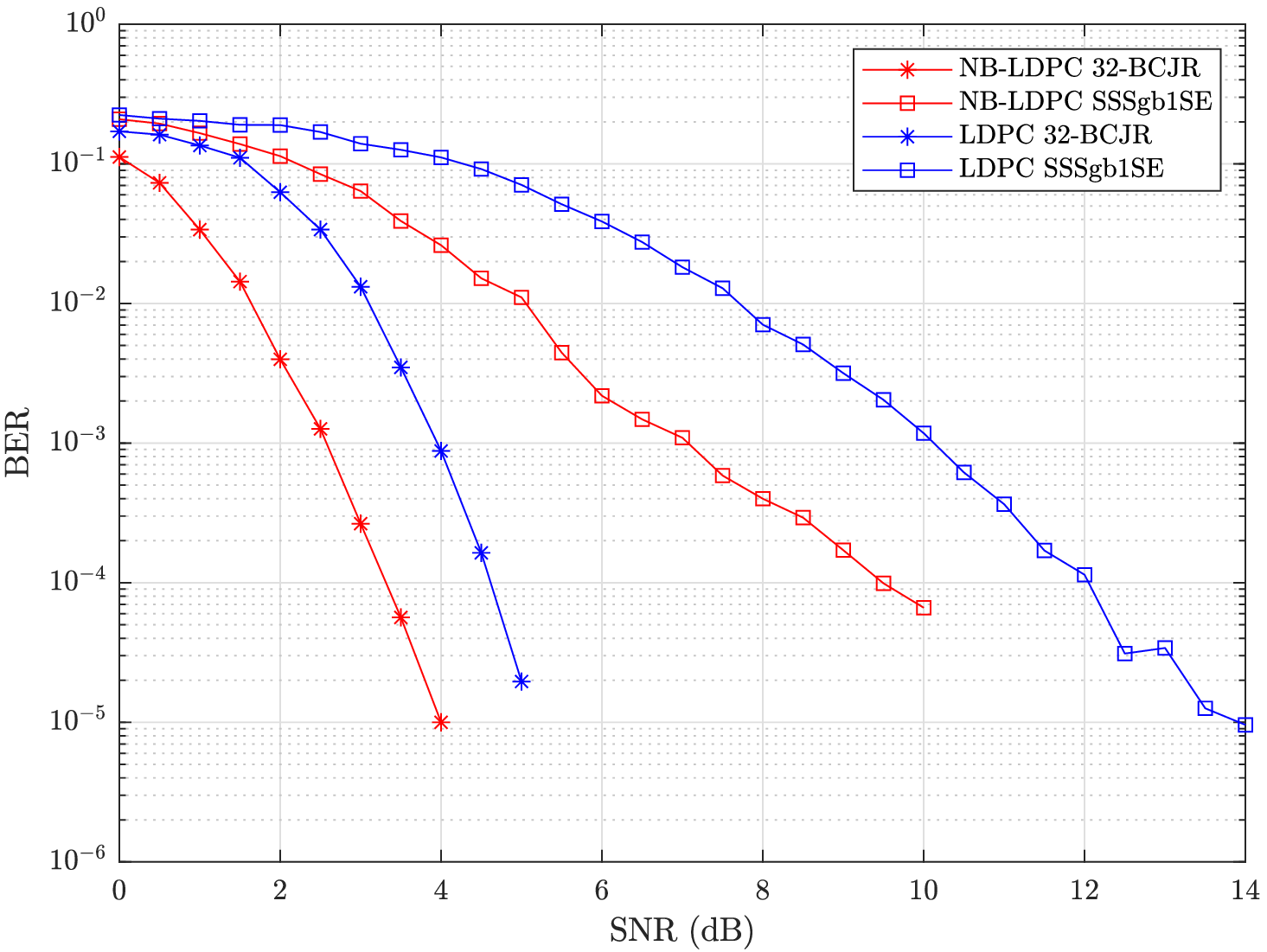}}
		\hfill
	\subfloat[\label{ber307}]{
		\includegraphics[width=0.3\textwidth]{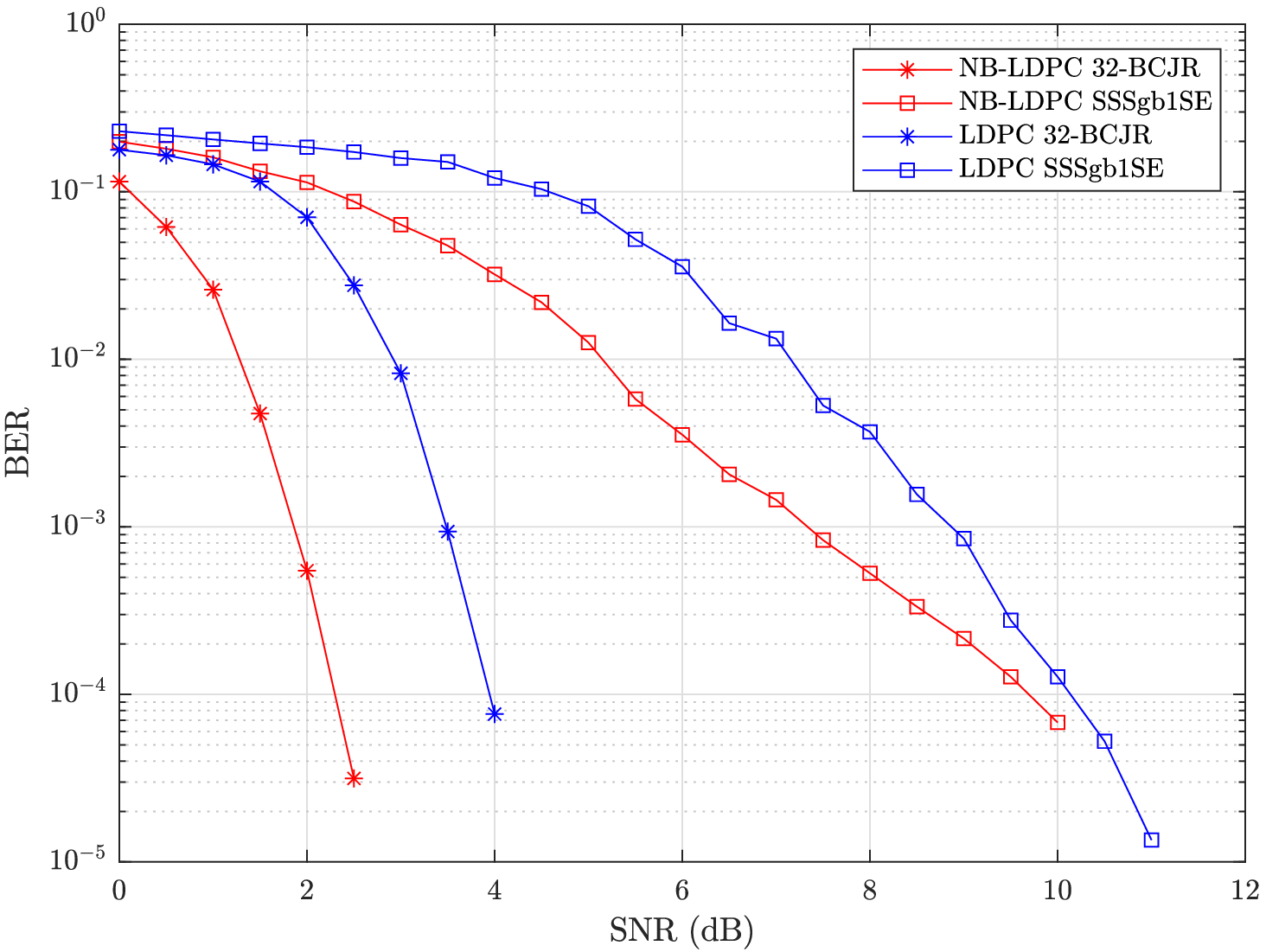}}
		\hfill
	\caption{BER performances of LDPC and NB-LDPC with SSSgb1SE and 32-BCJR FTN signaling systems for $\tau=0.7$ with (a) $N=128$ for LDPC and $N = 120$ for NB-LDPC, (b) $N=256$ for LDPC and $N=264$ for NB-LDPC and (c) $N=512$ for LDPC and $N=504$ for NB-LDPC.}
	\label{berfig07}
\end{figure*}

A message vector $\pmb{R}_i$ and a location vector $\pmb{\beta}_{\pmb{R}_i}$ form a 2-dimensional permutation matrix. $\pmb{\beta}_{\pmb{R}_i}$ needs to be permuted in V2C stage. The new location vector is found as $\pmb{\beta}_{\pmb{Q}_i}=H_{i,j}\circ\pmb{\beta}_{\pmb{R}_i}$. Hence, the original location vector needs to be found in C2V stage. An inverse permutation is applied as $\pmb{\beta}_{\pmb{R}_i}=(H_{i,j}\circ\pmb{\beta}_{\pmb{Q}_i})^{-1}$.

Having a set of messages, $S$, connected to check node, a configuration set $\mathcal{L}(\beta_{\pmb{A}} [i'])$ is a list of combination of elements in $GF(q)$, such that $\sum_{\pmb{\beta}_{\pmb{k}}\in S\setminus \pmb{\beta}_{\pmb{A}},i\in GF(q)\setminus \alpha_{i'-1}} H_{i,j} \beta_{\pmb{k}} [i]=H_{i',j} \beta_{\pmb{A}} [i']$. Check node processing was proposed as an adaptation of BP decoder to decode NB-LDPC codes in \cite{fastnbldpc14}. The method finds the sum of products of likelihood configurations that refers to the convolution of $d_c$ messages, where $d_c$ is the number of messages coming to a check node. C2V stage of EMS can be considered as an approximation of BP decoder and significantly reduces check node processing complexity. A scenario with a high number of $d_c$ and $q$, $q^{d_c}$ input configurations must be evaluated. ECN approach reduces this number by implementing elementary nodes having only two inputs.

Assuming inputs $\pmb{U}$ and $\pmb{I}$ having sorted $n_m$ elements decreasingly and $\pmb{V}$ as output, the index vectors $\beta_{\pmb{U}}$, $\beta_{\pmb{I}}$ and $\beta_{\pmb{V}}$ form the configuration set $\mathcal{L}(\beta_{{\pmb{V}}_i})$. Thus, $\beta_{\pmb{V}}[i]=\beta_{\pmb{U}} [j]\oplus \beta_{\pmb{I}} [k]$ needs to be satisfied. The resulting output vector $\pmb{V}$ becomes
\begin{align}
V[i]=\max_{\mathcal{L}(\beta_{{\pmb{V}}_i})} (U[j]+I[k]), \  j,k\in \{0, ..., n_m-1\}\setminus{i}.\label{eq14}
\end{align}
Equation \eqref{eq14} is simplified by applying L-bubble check algorithm \cite{boutillon15}. This adaptation introduces a further complexity reduction of $\sqrt{n_m}$. Thus, the vector $\pmb{Q}$ is found by using ECNs and permuting the indices back according to $\pmb{H}$. The syndrome is applied to the posterior probabilities and the decoder is halted if a valid codeword is found. The decoder gives $\hat{\pmb{c}}=\big[\begin{matrix}\hat{\pmb{p}} & \hat{\pmb{k}}\end{matrix}\big]$ as output upon finishing iterations, and consequently, $\hat{\pmb{k}}$ becomes the output message of NB-LDPC coded FTN signaling transmission.

\section{Simulation Results}

\begin{figure*}[t]
	\centering
	\subfloat[\label{ber108}]{
		\includegraphics[width=0.3\textwidth]{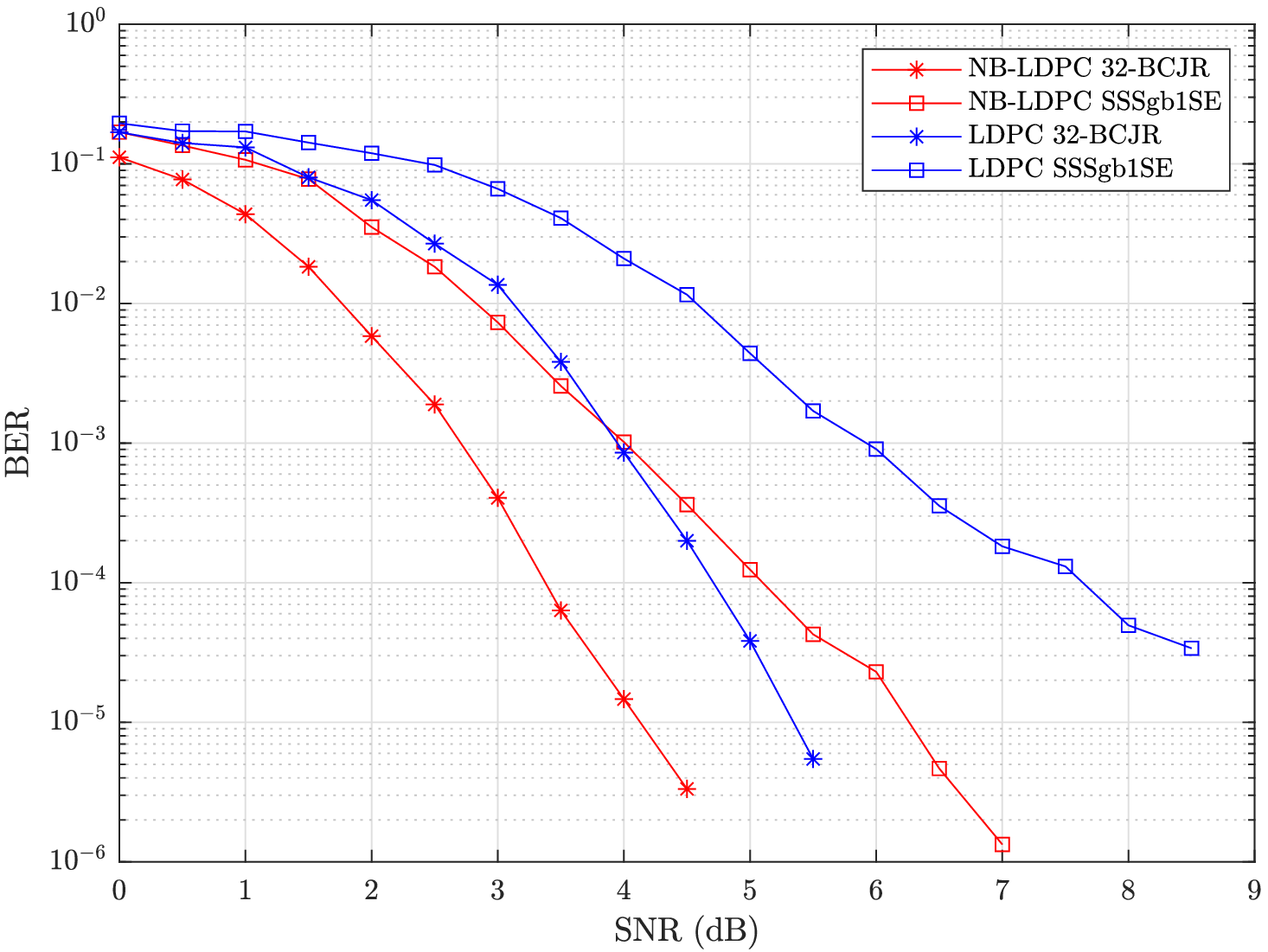}}
		\hfill
	\subfloat[\label{ber208}]{
		\includegraphics[width=0.3\textwidth]{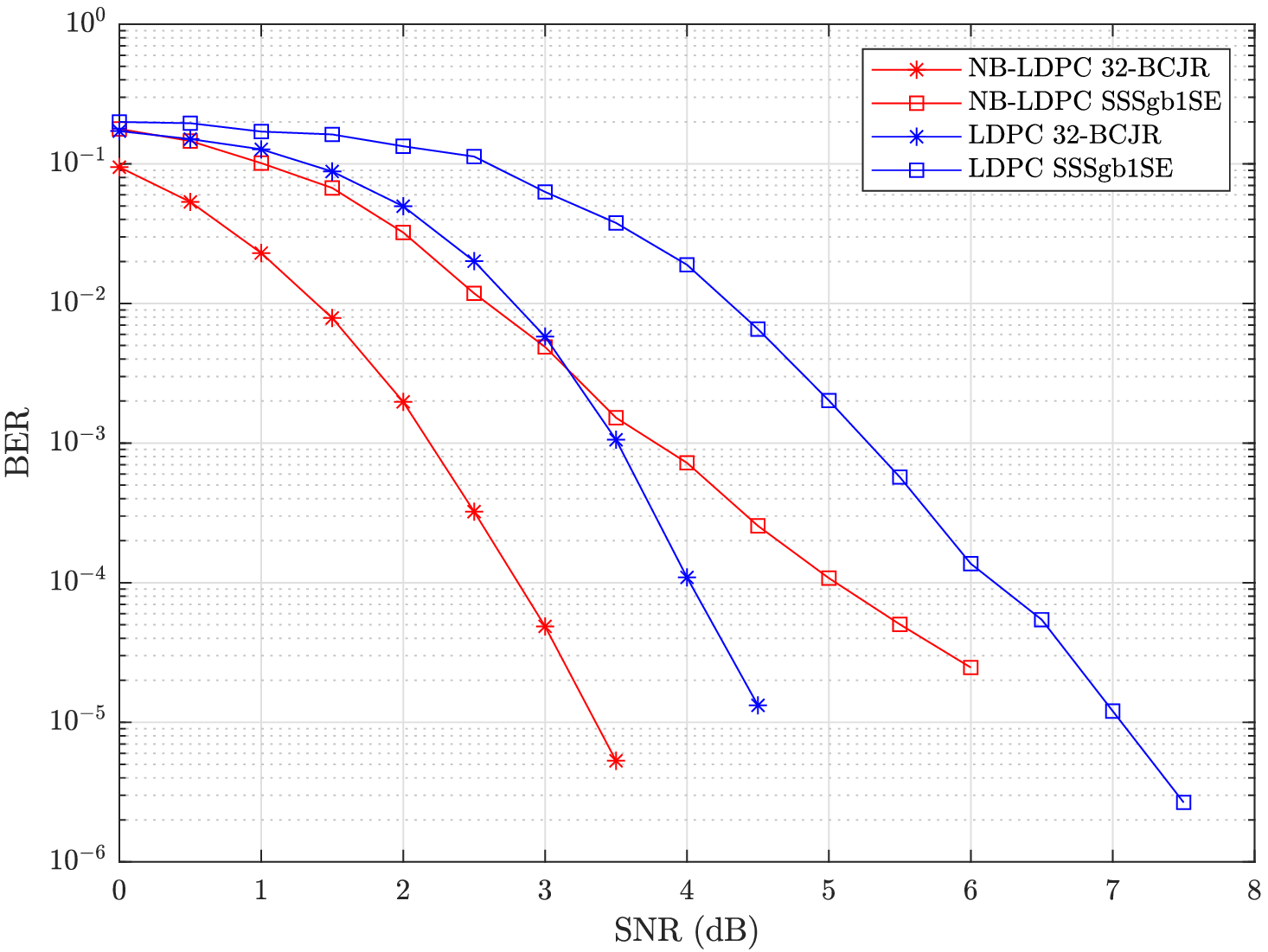}}
		\hfill
	\subfloat[\label{ber308}]{
		\includegraphics[width=0.3\textwidth]{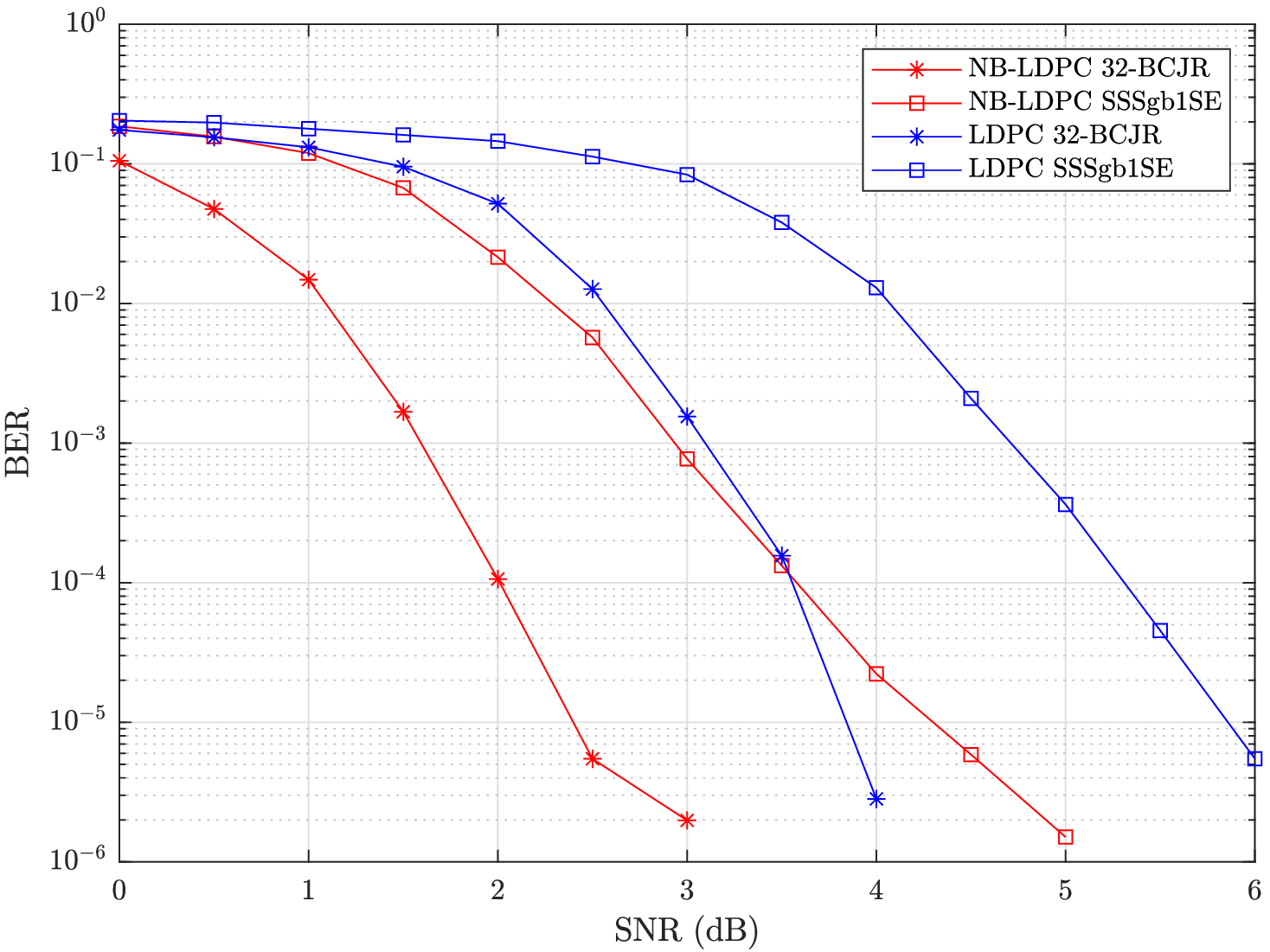}}
		\hfill
	\caption{BER performances of LDPC and NB-LDPC with SSSgb1SE and 32-BCJR FTN signaling systems for $\tau=0.8$ with (a) $N=128$ for LDPC and $N = 120$ for NB-LDPC, (b) $N=256$ for LDPC and $N=264$ for NB-LDPC and (c) $N=512$ for LDPC and $N=504$ for NB-LDPC.}
	\label{berfig08}
\end{figure*}

\begin{figure*}[t]
	\centering
	\subfloat[\label{ber109}]{
		\includegraphics[width=0.3\textwidth]{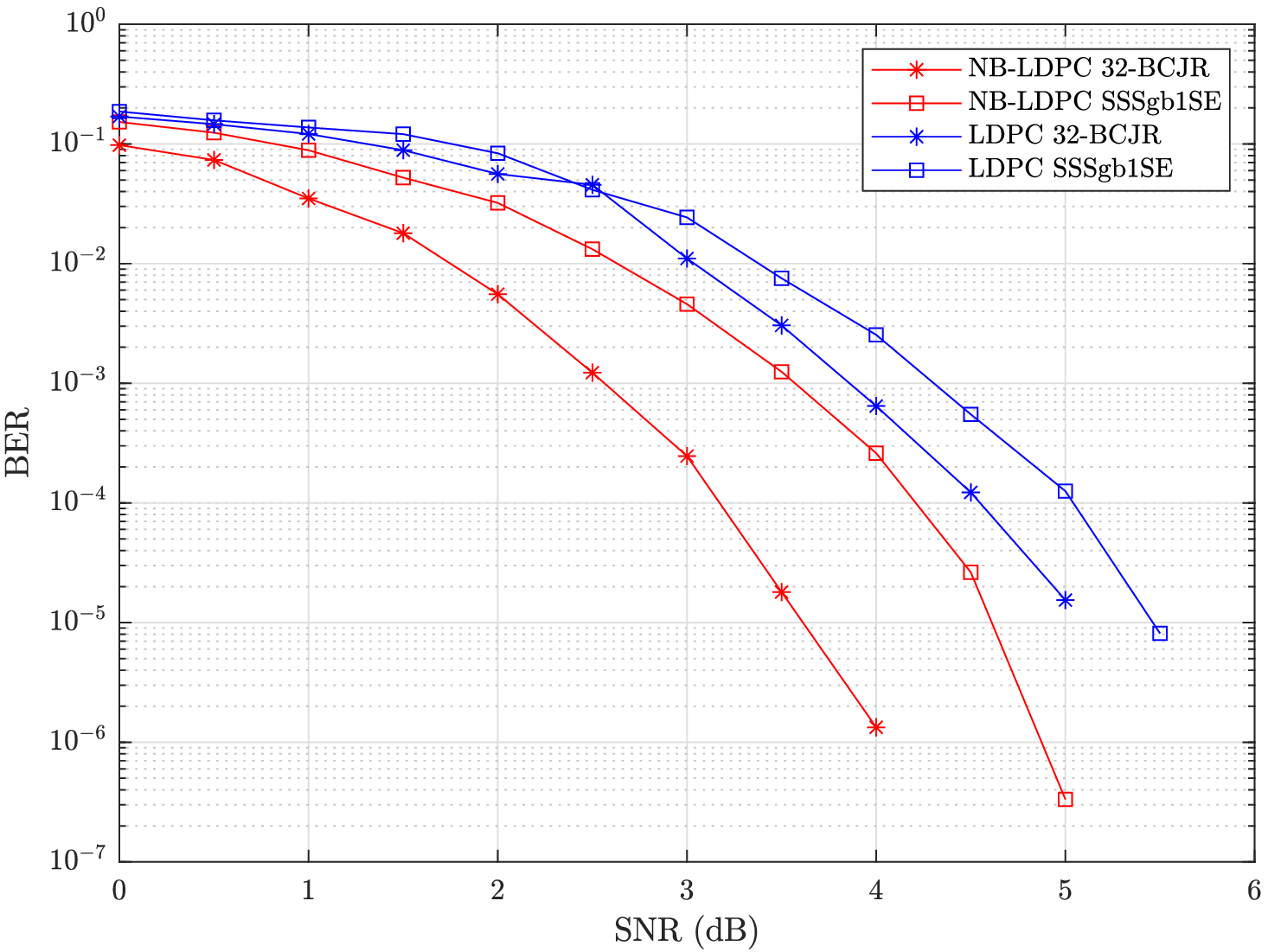}}
		\hfill
	\subfloat[\label{ber209}]{
		\includegraphics[width=0.3\textwidth]{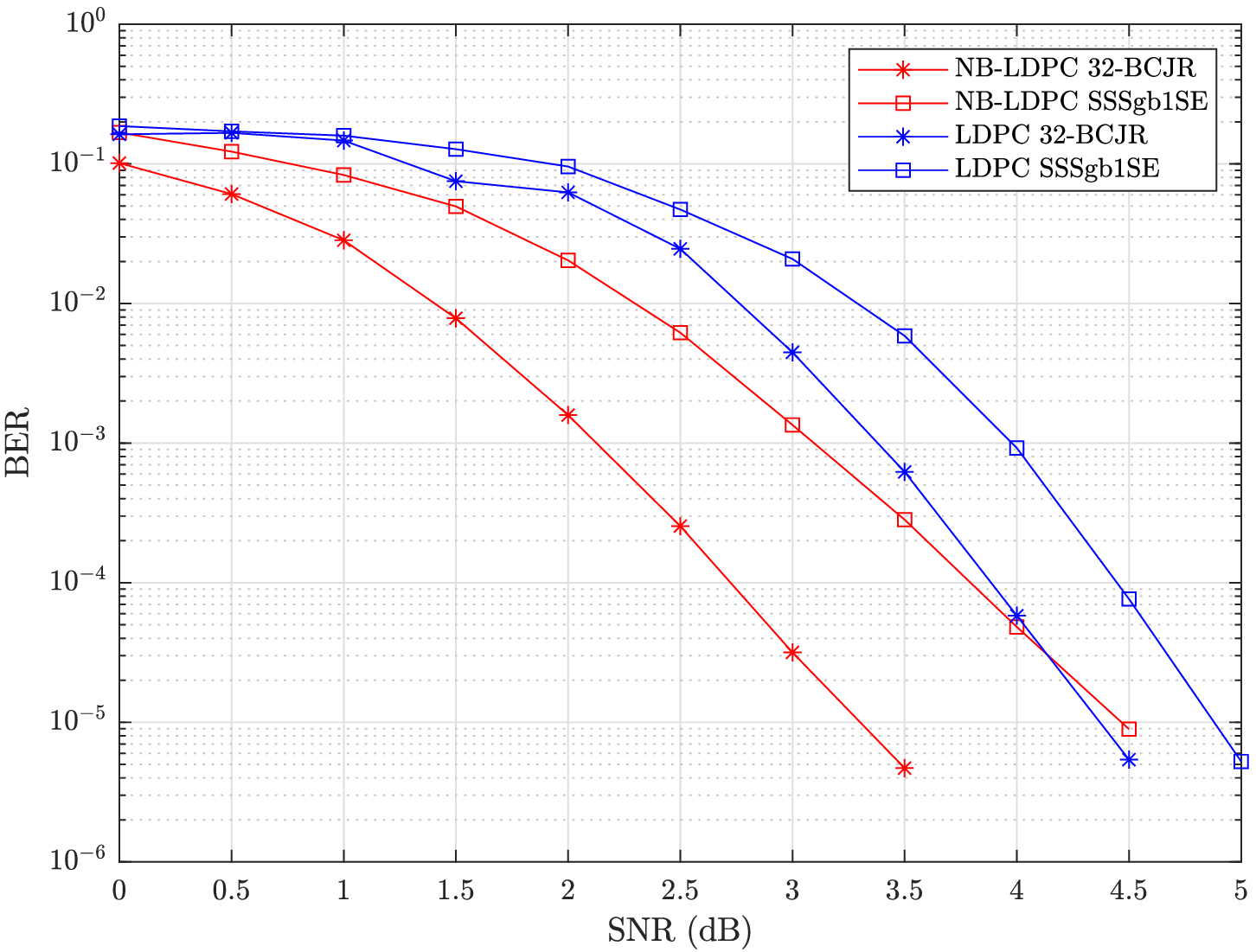}}
		\hfill
	\subfloat[\label{ber309}]{
		\includegraphics[width=0.3\textwidth]{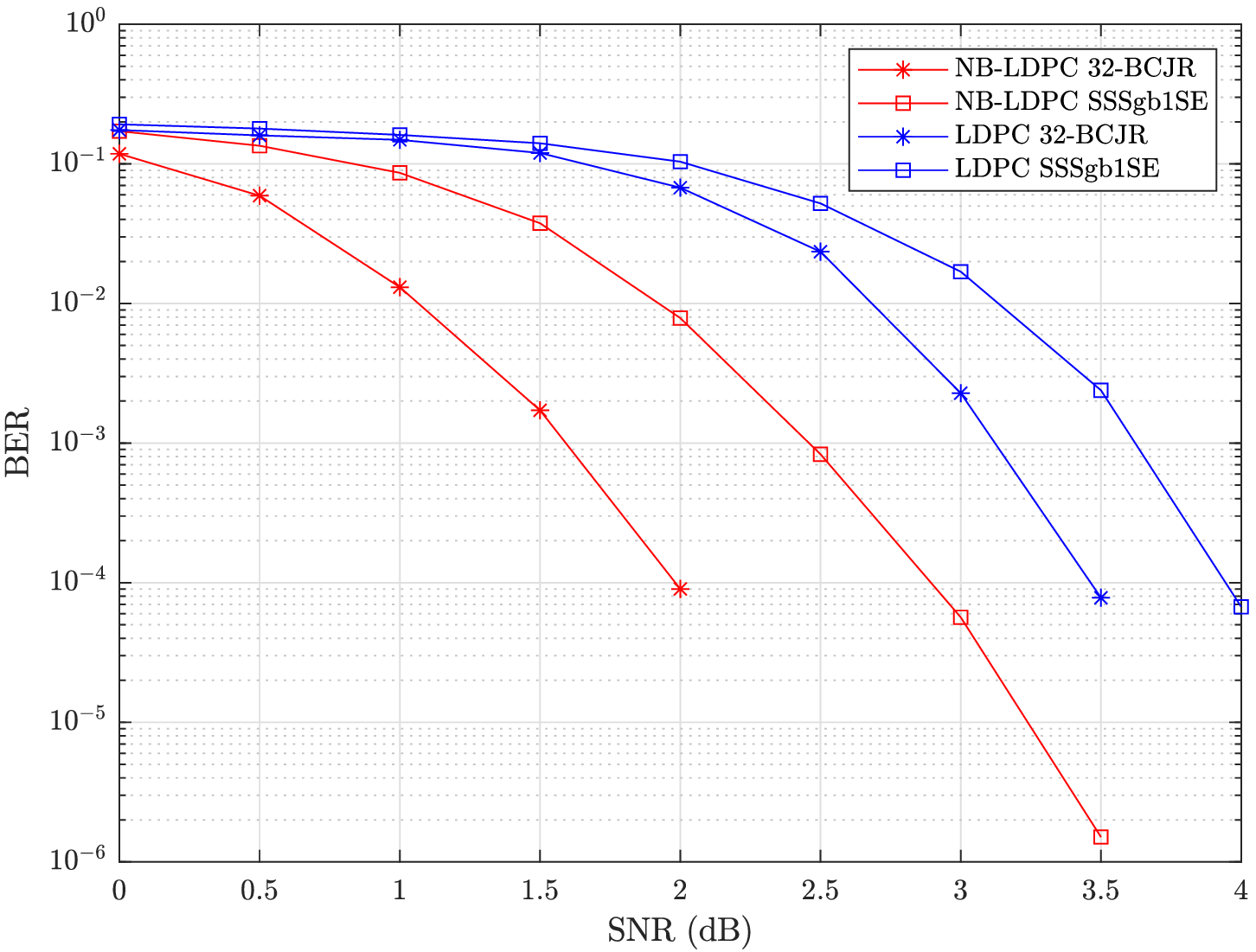}}
		\hfill
	\caption{BER performances of LDPC and NB-LDPC with SSSgb1SE and 32-BCJR FTN signaling systems for $\tau=0.9$ with (a) $N=128$ for LDPC and $N = 120$ for NB-LDPC, (b) $N=256$ for LDPC and $N=264$ for NB-LDPC and (c) $N=512$ for LDPC and $N=504$ for NB-LDPC.}
	\label{berfig09}
\end{figure*}
In this section, we investigate the performances of coded FTN signaling transmission for short packet communications for time acceleration parameters of 0.7, 0.8 and 0.9. As discussed earlier, two extreme FTN detectors are used: 32-BCJR as the close-to-optimal FTN signaling detector and SSSgb$1$SE as the lower-complexity FTN signaling detector. The BER performances of coded BPSK FTN signaling are investigated with both NB-LDPC and LDPC codes having code rates of 0.5. The codeword lengths of the LDPC codes are 128, 256, and 512 bits. The NB-LDPC codes were chosen in $GF(2^6)$, thereby each codeword symbol is represented with 6 binary digits. Considering that the code rate is 0.5; hence, the block lengths must be divisible by 12. For a fair comparison with LDPC, the NB-LDPC codes have codeword lengths of 120, 264 and 504 bits and automatically generated for $d_v=2$ \cite{web17}. The truncation size $n_m=20$, offset = 0.3, and 4-bubble check algorithm are used for the EMS decoder. NB-LDPC and LDPC decoders are set to find 100 frame errors in $5\times 10^5$ frames sent and 10 maximum iterations are used.

In Fig.~\ref{berfig07}, the BER performances of 32-BCJR and SSSgb1SE detection algorithms used with LDPC and NB-LDPC channel codes are depicted for $\tau=0.7$. As expected, the 32-BCJR FTN detector performs better under the same ISI levels compared to SSSgb1SE. At a BER of $10^{-4}$, the performance of LDPC with 32-BCJR ($N=128$) is 1.45 dB better compared to NB-LDPC with SSSgbKSE ($N=120$). For the same BER, LDPC with 32-BCJR detector surpasses NB-LDPC with SSSgb1SE with 1.23 dB and 1.44 dB for 256 and 512 bits of LDPC block lengths (264 and 504 for NB-LDPC), respectively. However, NB-LDPC with 32-BCJR detector still produces approximately 1 dB better performance compared to LDPC with 32-BCJR detector according to the SNR values that lead to a BER of $10^{-4}$ in Fig.~\ref{berfig07}.

Fig.~\ref{berfig08} shows BER performances of 32-BCJR and SSSgb1SE detection algorithms used with LDPC (128, 265 and 512 bits of block lengths) and NB-LDPC (120, 264 and 512 bits of block lengths) channel codes for the case $\tau=0.8$. In this case, NB-LDPC with SSSgb1SE having 120 bits of block length surpasses LDPC with 32-BCJR with 128 bits of block length until 3.8 dB SNR (Fig.~\ref{ber108}) and has about 0.3 dB performance gain at $10^{-4}$. The intersection points of BER curves shift to 3.4 dB and 3.5 dB SNR and have 0.24 dB and 0.3 dB performance gains for the LDPC codes having 256 and 512 of bits block lengths (264 and 504 bits for NB-LDPC), respectively. One can conclude from Fig.~\ref{berfig08} that at $\tau = 0.8$, it is not advised to use complex FTN signaling detectors and a simple FTN detector is sufficient.

On the other hand, for the case of $\tau=0.9$ in Fig.~\ref{berfig09}, NB-LDPC with SSSgb1SE detection outperforms LDPC with 32-BCJR detection. For a BER of $10^{-4}$, 0.57 dB, 0.11 dB and 0.35 dB performance gains are observed between 128, 256 and 512 bits of block lengths for LDPC and 120, 256 and 504 bits for NB-LDPC, respectively. The performance of NB-LDPC with SSSgb1SE having 504 bits of codeword length is always better than LDPC with 32-BCJR detector for 512 bits of codeword length for $\tau=0.9$. The performance gain varies between 0.53 and 0.87 dB. The performance of NB-LDPC with SSSgb1SE is also better for the 120 bits of block length case and the gain changes between 0.3 and 0.63 dB, compared to LDPC with 32-BCJR with 128 bits of block length. However, Fig.~\ref{ber209} shows that the intersection for codewords having lengths of 256 (for LDPC) and 264 (for NB-LDPC) bits occurs at 4.12 dB SNR yielding a BER of $3.6\times 10^{-4}$. One can see from Fig.~\ref{berfig09} that for high values of $\tau$, there is no benefits from using high complexity FTN signaling detectors with LDPC codes, and a low-complexity FTN signaling detector is sufficient with NB-LDPC codes.

\section{Complexity Evaluation}

\begin{figure*}[t]
	\centering
	\subfloat[\label{noo128}]{
		\includegraphics[width=0.3\textwidth]{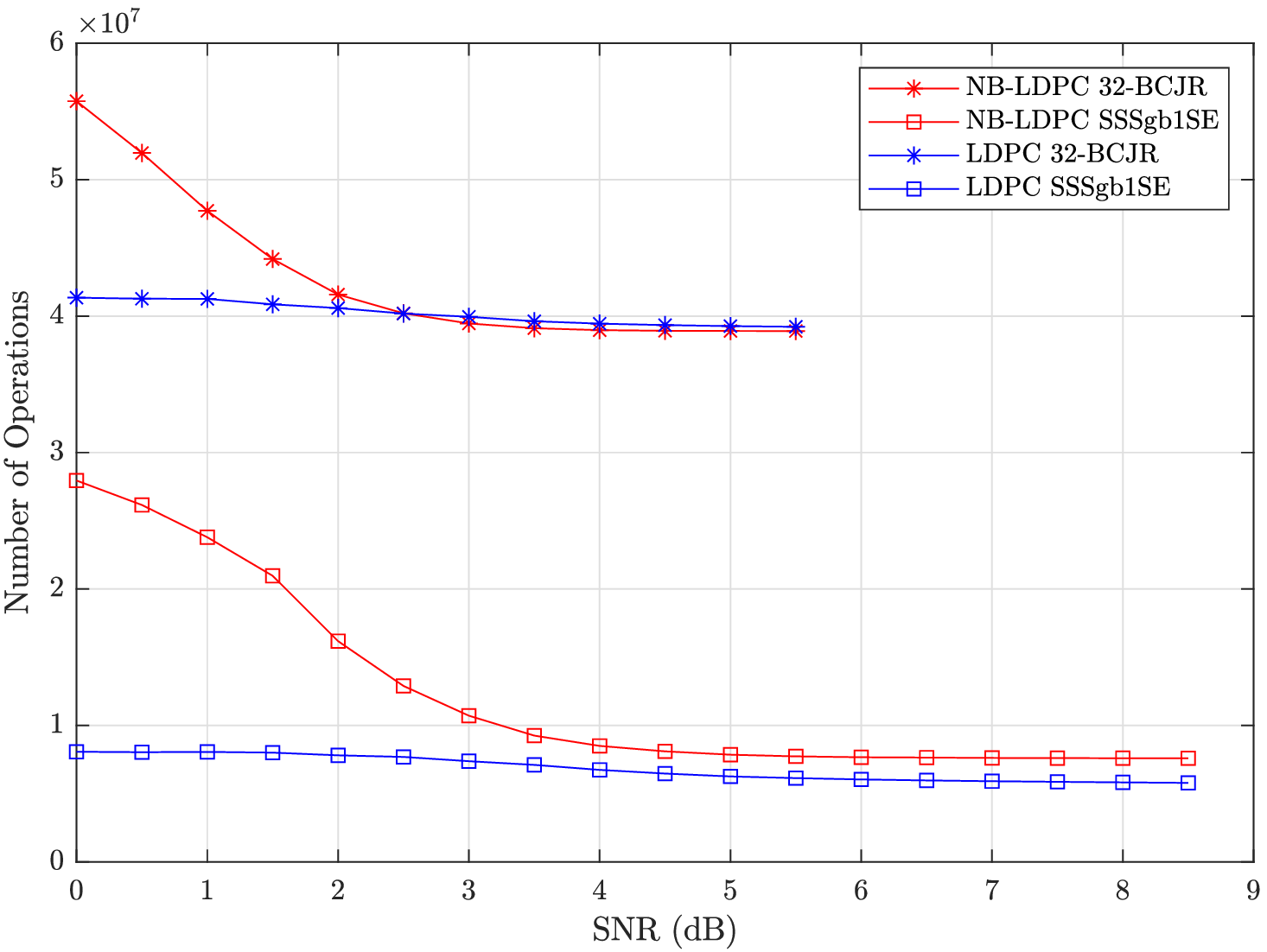}}
		\hfill
	\subfloat[\label{noo256}]{
		\includegraphics[width=0.3\textwidth]{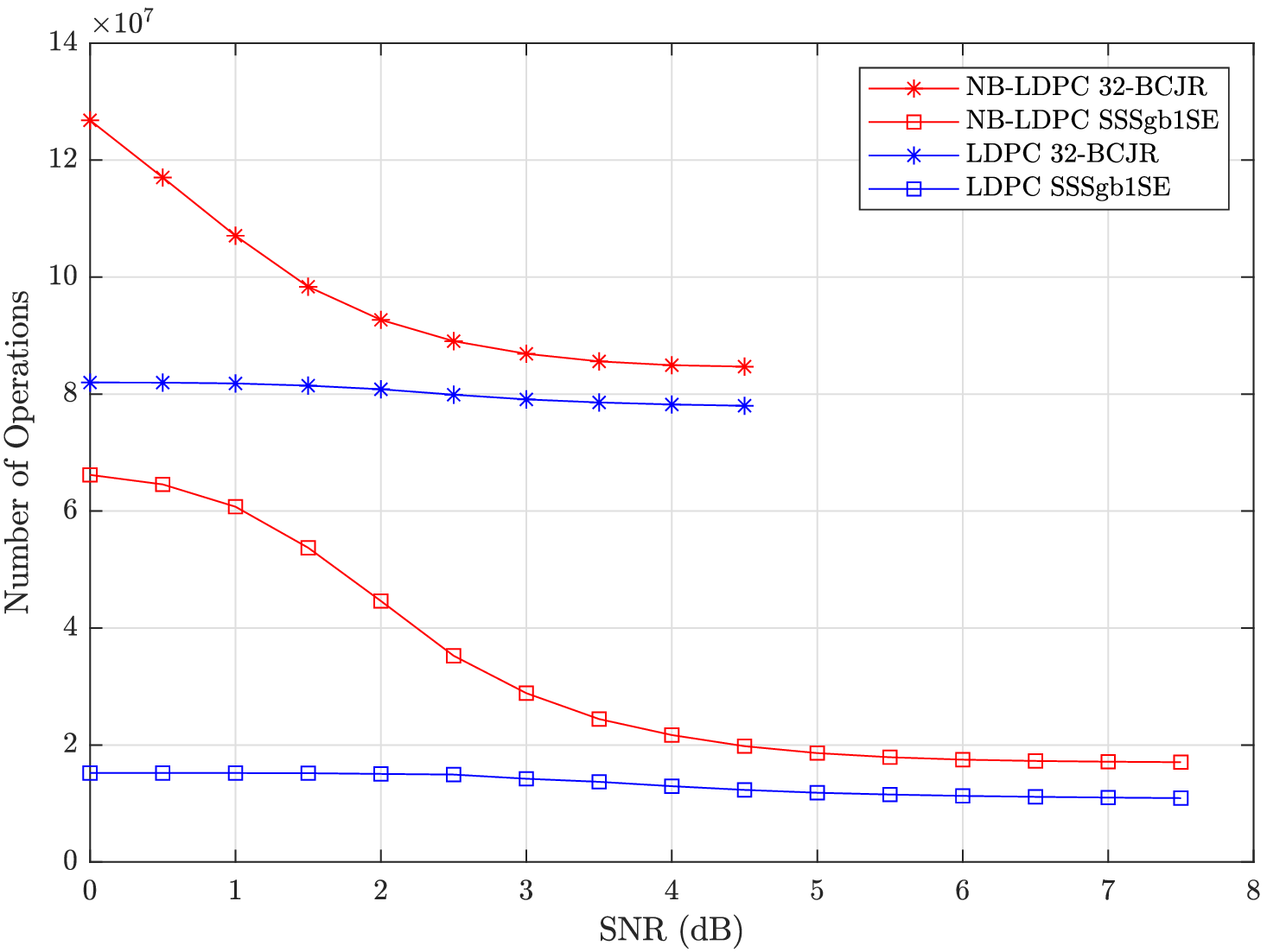}}
		\hfill
	\subfloat[\label{noo512}]{
		\includegraphics[width=0.3\textwidth]{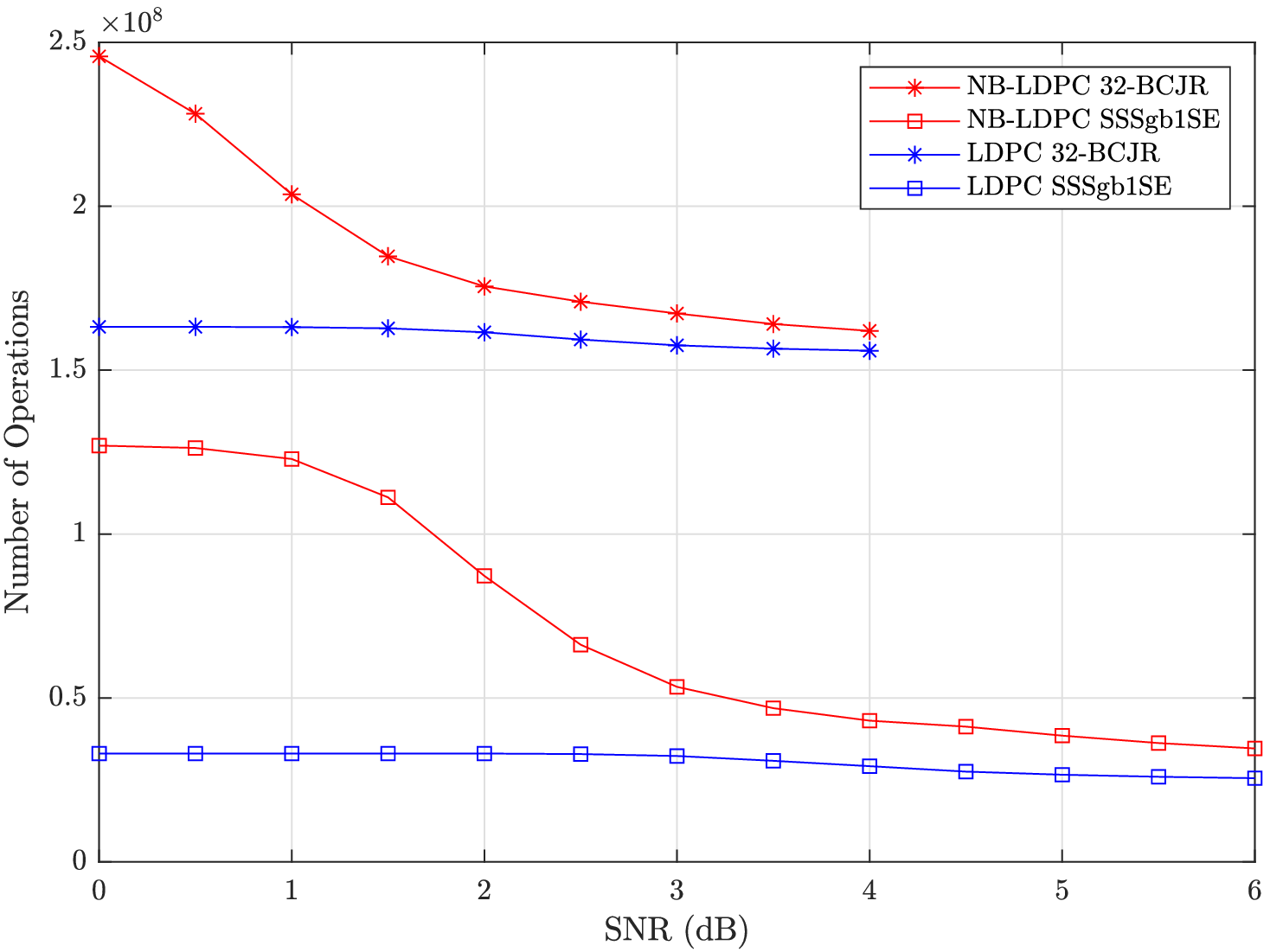}}
		\hfill
	\caption{Total number of operations with 10-bit numbers for codeword lengths at $\tau = 0.8$ (a) $N=128$ for LDPC and $N = 120$ for NB-LDPC, (b) $N=256$ for LDPC and $N=264$ for NB-LDPC and (c) $N=512$ for LDPC and $N=504$ for NB-LDPC.}
	\label{noofig}
\end{figure*}

The number of operations are presented in Table \ref{nootab} in terms of $d_c$, $d_v$, and truncation size $n_m$. The definitions in Table \ref{nootab} are based on C2V, V2C, and syndrome control stages combined. The min-sum algorithm (MSA) for LDPC requires $d_c$ additions and $d_c$ subtractions with real numbers. However, the EMS algorithm has $n_m (9d_c+8d_v-22)$ real additions including L-bubble check algorithm. The syndrome check and search operations contribute to $2d_v-3$ comparisons and $d_v(d_v-1)$ XOR operations in MSA. The definition $n_m \log_2 n_m$ in NB-LDPC comparisons refers to search in the L-bubble check algorithm, while the remaining part is the maximum and sorting operations. The XOR operations for NB-LDPC EMS algorithm are the combination of additions in finite field arithmetic and syndrome check stage.

\begin{table}[t]
	\caption{Number of operations per iteration.}
	\begin{center}
		\begin{tabular}{|c|c|c|}
			\hline
			\text{ }&\multicolumn{2}{|c|}{\textbf{Channel Codes}} \\
			\cline{2-3} 
			\text{ } & {{LDPC}}& 					
			{{NB-LDPC}}\\
			\hline
			Addition 		& $d_c$ 			& $n_m (9d_c+8d_v-22)$ \\
			\hline
			Subtraction 	& $d_c$ 			& $0$ \\
			\hline
			Comparison$^{\mathrm{a}}$ 	& $2 d_v-3$ 		& $x(6d_c+4d_v-18)+n_m(4d_v-6)-2$ \\
			\hline
			XOR				& $d_v^2-d_v$ 	& $(n_m  \log_2 q) (9d_c-18)$ \\
			\hline
			\multicolumn{3}{l}{$^{\mathrm{a}}$ $x=n_m\log_2 n_m.$}
		\end{tabular}
		\label{nootab}
	\end{center}
\end{table}

Fig.~\ref{noofig} shows the total number of operations {for $\tau=0.8$}, reduced to AND gate level. The total number of operations is calculated with the help of Table \ref{nootab} by multiplying the average number of iterations that comes from our simulations and the number of operations given in Table \ref{nootab}. The hardware implementation of the real numbers in Table \ref{nootab} is assumed to have 10-bits. The effects of FTN detectors and decoder iterations are also taken into consideration. The sharp decrease of the number of iterations of NB-LDPC codes lowers the complexity considerably compared to LDPC with the same FTN detection algorithm for each case. As can be seen from Fig.~\ref{noofig}, the complexity of LDPC and 32-BCJR is 3 to 5 times higher when compared to NB-LDPC and SSSgb1SE. Since the starting instances when LDPC and 32-BCJR performing better than NB-LDPC and SSSgb1SE lie at SNR values between 3 and 4 dB, the differences of total number of operations for the same SNR values conclude that using NB-LDPC and SSSgb1SE benefits the system both by having a slightly better performance and requiring a lower number of operations. Although the average number of iterations changes for different time acceleration parameters, the ratio of the total number of operations of the LDPC and 32-BCJR versus the NB-LDPC and SSSgb1SE will be similar to the ratio in Fig.~\ref{noofig} at $\tau = 0.8$ for sufficiently high SNRs.

\section{Conclusion}
FTN signaling offers promising improvements of the SE of next-generation URLLC systems. The adoption of reduced-complexity FTN signaling detectors in short packet communications requires compatible error correction codes. We investigated if the low-complexity SSSgb\textit{K}SE FTN signaling detector combined with NB-LDPC codes for short packet communications outperforms the use of close-to-optimal and more complex FTN signaling detectors, e.g., M-BCJR, with LDPC codes. Our investigation showed that NB-LDPC with SSSgb\textit{K}SE FTN signaling detection has 0.3 dB better BER performance than LDPC with M-BCJR FTN signaling detection until around 3.5 dB SNR for $\tau=0.8$ for short block lengths. Furthermore, 0.35 and 0.57 dB gains were achieved for NB-LDPC with SSSgb\textit{K}SE FTN signaling detection for 128 and 512 bits of codewords for $\tau=0.9$. The complexity evaluations indicated that the complexity of NB-LDPC with SSSgb\textit{K}SE FTN signaling detection on short block lengths is lower than its counterpart of LDPC with M-BCJR FTN signaling detection.

\end{document}